\documentclass[a4paper,fleqn]{cas-sc}
\makeatletter
\@fleqnfalse
\makeatother

\usepackage[numbers]{natbib}

\def\tsc#1{\csdef{#1}{\textsc{\lowercase{#1}}\xspace}}
\tsc{WGM}
\tsc{QE}
\tsc{EP}
\tsc{PMS}
\tsc{BEC}
\tsc{DE}

\usepackage{tikz}
\usetikzlibrary{shapes.geometric, positioning, arrows}  
\usepackage{float}

\tikzstyle{block} = [rectangle, draw, fill=white, 
text width=9em, text centered, rounded corners, minimum height=4em]
\tikzstyle{line} = [draw, -latex]  

\usepackage{algorithm2e}
\usepackage{algpseudocode}

\usepackage{amsmath} 

\SetAlCapSkip{1em}

\SetKwInput{KwInput}{Input}                
\SetKwInput{KwOutput}{Output}  
\SetKwFunction{FMain}{MOM}
\SetKwFunction{FMainn}{GAN}
\SetKwFunction{FTest}{Branching}
\SetKwProg{Fn}{Function}{:}{end}

\begin{document}

\let\WriteBookmarks\relax
\def\floatpagepagefraction{1}
\def\textpagefraction{.001}

\title [mode = title]{                   
Markov processes for enhanced deepfake generation and detection}  

\author[1]{Michael A. Kouritzin}


\ead{michaelk@ualberta.ca}

\affiliation[1]{{Department of Mathematical and Statistical Sciences, University of Alberta}}

\author[2]{Ian Zhang}
\cormark[1]
\affiliation[2]{{Department of Statistical Sciences, University of Toronto}}
\ead{ianz.zhang@mail.utoronto.ca}
\author[3]{Jyoti Bhadana}
\affiliation[3]{{Department of Mathematics, University of Texas at Arlington}}
\ead{jyoti.bhadana@uta.edu}

\author[1]{Seoyeon Park}

\ead{sp8@ualberta.ca}

\cortext[cor1]{Corresponding author}

\begin{abstract}
New and existing methods for generating, and especially detecting, deepfakes are investigated and compared on the simple problem of authenticating coin flip data. 
Importantly, an alternative approach to deepfake generation and detection, which uses a Markov Observation Model (MOM) is introduced and compared on detection ability to the traditional Generative Adversarial Network (GAN) approach as well as Support Vector Machine (SVM), Branching Particle Filtering (BPF) and human alternatives.
MOM was also compared on generative and discrimination ability to GAN, filtering and humans (as SVM does not have generative ability).
Humans are shown to perform the worst, followed in order by GAN, SVM, BPF and MOM, which was the best at the detection of deepfakes. 
Unsurprisingly, the order was maintained on the generation problem with removal of SVM as it does not have generation ability.
\end{abstract}


\maketitle

\section{Introduction}
Fake or synthetic content generated though advanced deep learning techniques that appears authentic in the eyes of a human being is called a “deepfake”. The most common form of deepfakes involves the generation and manipulation of human imagery. Deepfake technology has creative and productive applications in entertainment, education, content creation, computer vision, natural language processing, and human-level control \cite{Westerlund}. Deepfakes have spread to other domains and media such as forensics, finance, and healthcare \cite{Taulli}. However, deepfakes also pose substantial risks such as misinformation, privacy invasion, and identity theft. While they are usually trained to foil discriminators, the objective of the generated content is to fool a human, not a machine, and for that reason they have garnered heightened attention. Still, a machine can also be a valuable tool in its detection.

Rather than studying full audio-visual deepfakes directly, this paper focuses on a controlled coin-flip sequence setting. This lets us compare different detection and generation methods under clear data-generating assumptions, while still keeping the core fake-versus-real classification problem.

This problem is related to earlier work on distinguishing truly random sequences from fake sequences produced by humans or machines \cite{Kouritzin,warrenReexaminationBiasHuman2018,koeveringHowRandomRandom2024}. It is also connected to probabilistic sequence models such as hidden Markov models and pairwise Markov chains, which provide natural tools for sequence generation, filtering, and likelihood-based classification \cite{rabinerTutorialHiddenMarkov1989,pieczynskiPairwiseMarkovChains2003}.

In this paper, we compare several approaches in this controlled sequence setting, including GAN, SVM, branching particle filtering (BPF), and the proposed Markov Observation Model (MOM). MOM replaces the neural-network generator/discriminator framework with a latent probabilistic sequence model that supports both sequence generation and likelihood-based classification. This allows us to study the trade-off between probabilistic structure, detection performance, and computational cost in a setting where the assumptions are explicit and the methods are directly comparable.

Consequently, we will introduce a probabilistic GAN technology.
For centuries, mathematicians have expressed interest in the classification problem, to distinguish between fake data "randomly generated" by humans and real data produced by random physical processes. 
A natural extension to this problem is to distinguish between deepfakes
"randomly generated" by computers and real data. 

There is a long-standing statistical perspective for analyzing sequential data through probabilistic sequence models, including Markov-chain and hidden Markov model (HMM) frameworks. These models support likelihood-based inference, recursive decoding/inference algorithms, and interpretable parametrization, and they are often studied together with model-selection tools such as information criteria and likelihood-ratio based procedures \cite{costaModelSelectionHidden2010, rabinerTutorialHiddenMarkov1989}. Our Markov Observation Model (MOM) relates to this, and is also related to broader generalizations such as pairwise Markov-chain models \cite{pieczynskiPairwiseMarkovChains2003}; it provides a generative mechanism for sequence synthesis and a likelihood-based route for classification.

At the same time, our paper focuses on a controlled sequence-based authentication setting with binary observations, where probabilistic sequence models and learning-based baselines can be compared under transparent data-generating assumptions. Thus, our contribution is a controlled and interpretable benchmark setting, rather than a direct multimodal audio--visual deepfake benchmark \cite{rosslerFaceForensicsLearningDetect2019,licelebdf2020}.

While many deepfake detectors focus on artifact cues or deep neural architectures for image/video data \cite{afcharMesoNetCompactFacial2018, liFaceXRayMore2020,rosslerFaceForensicsLearningDetect2019}, our goal here is to study a controlled sequence-based authentication setting where probabilistic and learning-based approaches can be compared under transparent data-generating assumptions. This also complements recent deepfake-detection literature emphasizing reliability challenges such as transferability and robustness \cite{wangDeepfakeDetectionComprehensive2022}.

In this paper, we study a controlled coin-flip authentication problem and propose a simple method that simultaneously simulates and classifies sequences as real or (deep) fake using a modified probabilistic approach to traditional generative adversarial networks (GANs). Many researchers have explored the classification of coin-flip sequences as real or fake \cite{Revesz}, \cite{Schilling}, whereas others have developed algorithms for generating fake flips whilst capturing the properties of a real flip \cite{Kouritzin}. Another example of this problem proposed by Green \cite{Batanero} was addressed to school aged children, with the task being that they needed to correctly identify which sequences were "real" or "random", along with which sequences appeared to be "produced by a fair coin" or "fake" \cite{Renelle}. This problem was then used to understand and model the "fake" behaviour. The power in the existing probabilistic approach comes from a separate hidden model, the signal model, that represents both the real and fake random behaviours, while the variables of interest, in this instance the coin flips, are modelled as the observations \cite{Kouritzin} that depend upon this signal. Generation then comes from signal, observation simulation. Detection, on the other hand, becomes estimating which behaviour is present in the signal, which can be done optimally by filtering theory \cite{Xiong} and practically by particle
filters \cite{Moral,Douc,Hol,Murray}. This existing approach entails encoding all fake and real models into the same signal, usually by including some state variable that indicates which behaviour is active. The problems with expanding this approach beyond simple authentication problems like coin flips is that one needs to: i) have models for every type of fake as well as real behaviour, ii) amalgamate these models together into one signal model and iii) suffer extra processing costs by working with all behaviours simultaneously when they are amalgamated into one signal. This does not seem practical for large problems like authenticating real video content. We will suggest a new probabilistic approach below that both outperforms this existing approach and is more expandable to larger problems. The first step of the new approach is separating out the signal and employing recursive likelihood model selection, which we will also use herein with branching particle filters (BPFs). The recursive likelihood approach was introduced in \cite{kouritzinMarkovObservationModels2025}. BPFs, where branching is used in place of resampling to redistribute particles, were introduced by Crisan and Lyons \cite{Crisan} but these early algorithms suffered from extreme particle swings, which affected performance dramatically. \cite{Ballantyne} developed a stabler BPF.
However, it was only in \cite{Kouritzin1, Kouritzin4} that a fully stable algorithm was given where it could be shown that the number of particles in a BPF were controllable.
It was also shown there that BPFs can have recursive likelihood model selection capabilities with effective resampling.  
Finally, it was also shown that by reducing particle number fluctuations one improves performance and reliability \cite{Kouritzin1}.

To address these limitations, we introduce a new GAN-like architecture that utilizes a model and algorithm recently proposed by \cite{kouritzinMarkovObservationModels2025}. 
In particular, fake sequences are modelled as Markov chains with hidden states. This Markov Observation Model (MOM) replaces the neural networks within the traditional GAN architecture. MOM is simply a pairwise Markov chain $\begin{pmatrix}
    X & Y
\end{pmatrix}'$ with one step transition probabilities $p$ and initial distribution $\mu$, where one component of the chain $X$ is hidden and the other $Y$ is observable.
It is shown in \cite{kouritzinMarkovObservationModels2025} that a Baum-Welch-like forward backward Expectation-Maximization (EM) algorithm for both $p$ and $\mu$ holds in this setting and converges to (at least) local maxima of the likelihood function.
Like for HMM the forward-backward reduction of the computations produces a highly efficient learning algorithm. 
The \emph{generation} part of MOM comes, after the learning is complete, by simulating the pairwise Markov chain and then throwing the hidden part away.
The \emph{discrimination} part of MOM comes, after learning all possible (real and fake) models, by likelihood model selection, which is also shown in \cite{kouritzinMarkovObservationModels2025}, to determine if a given sequence is better represented by real or fake models. 

We chose to introduce our method on the simple coin flip problem in order
to compare it to many other methods and due to the availability of computer
resources.
However, there is no need to encode mathematical models into a signal
and there are efficient forward-backward algorithms for learning as well
as classifying.
Hence, there appears to be no reason why the MOM-based method can not be expanded to more practical and pressing deepfake detection problems in the future.

Our work is also related to a broader line of research on robust multimodal learning and robustness-aware generation/evaluation (e.g., in emotion recognition, multimodal fusion, and other perception tasks) \cite{baltrusaitisMultimodalMachineLearning2019,yangDefendingMultimodalFusion2021}. These works emphasize robustness to noise, perturbations, and modality shifts, which aligns with the broader motivation of reliable fake/real discrimination \cite{yangDefendingMultimodalFusion2021}. At the same time, our paper focuses on a controlled setting with binary observations, where probabilistic sequence models and learning-based baselines can be compared under transparent data-generating assumptions. We therefore present our method as a simple and interpretable test setting, rather than a full multimodal audio-visual deepfake benchmark \cite{licelebdf2020,rosslerFaceForensicsLearningDetect2019}.

Our work is also related to broader research on robust multimodal learning, including work on multimodal fusion, modality balancing, robustness to missing or noisy modalities, and lightweight or privacy-aware multimodal methods \cite{baltrusaitisMultimodalMachineLearning2019,yangDefendingMultimodalFusion2021,zhaoTEMPOTrainingtimeEquilibration2026,linWiViUFUnifiedFeature2026,tianFedKTNatureInspiredBayesian2026,zhuRMERDTRobustMultimodal2025,xiangIntegratingAudioVisual2026}. These papers show the broader importance of reliable discrimination in difficult settings. However, they study much richer multimodal problems than the one considered here. Our paper focuses instead on a controlled binary-sequence benchmark, which we use as a simple and interpretable test setting rather than a full multimodal audio-visual deepfake benchmark \cite{licelebdf2020,rosslerFaceForensicsLearningDetect2019}.

\subsection{Layout}
This paper is mainly divide into two parts, one is deepfake generation and the another one is deepfake detection. Specifically, in Section 2, we discuss methods of simulating a fake coin flip sequence; in Section 3, we compare the Markov Observation Model results to the alternative generative methods; in Section 4, we present our conclusions and future work. 
\subsection{Computer System}
All simulations were done on an \emph{M3 Pro MacBook Pro} in Python 3.10. In our study, we utilized a variety of packages and libraries to conduct simulations and analyses. The core libraries included NumPy for numerical computations, Pandas for data handling and analysis, and the random module for generating random numbers. We employed the Keras library to build and train the GAN model, leveraging layers such as Dense, LeakyReLU, BatchNormalization, and optimizers like Adam. We used scikit-learn, implementing models like Support Vector Classification (SVC) and One-Class SVM, and performed tasks such as cross-validation, and train-test splitting. Additionally, we assessed model performance using metrics from scikit-learn, including precision, recall, classification reports, ROC-AUC, average precision scores, and confusion matrices. These tools were integral to processing data, building models, and evaluating outcomes within our simulations.

\section{Methods of Data Generation}

This section describes the different ways coin flip sequences were created. 
We will use five sources: real (by random number generator), human fake,
the Simulator algorithm that can be thought of as the generator part of the 
BPF approach, the generative part of GAN and the generative part of MOM.
Generally, some initial data is created representing the two types, real and (deep) fake, to distinguish. Then, adversarial methods are used to create more deepfake data. Some initial deepfake data was produced by a Bernoulli Generation algorithm from \cite{Kouritzin}, termed the \emph{Simulator} here, capable of producing desired correlations to prior samples. Other deepfake data was created using Generative Adversarial Networks (GAN), and the newly proposed Markov Observation Model (MOM) generation. 

\subsection{Initial Data}

Adversarial networks require some initial training data or else some other means of setting up initial generator and discriminator rules.
We started the system with real, fake and deepfake data sequences as follows.

\subsubsection{Real Sequences}

We used Python's \texttt{random} package on an M3 Pro MacBook Pro to generate independent sequences of independent coin flip data.
We treated these as if they were real coin flips, even though they were computer generated.
For representation purposes, ones were interpreted as heads, and zeroes were interpreted as tails.

\subsubsection{Real Fakes}
We employed a group 15 students to create 137 sequences of 200 \emph{real fake} coin flips that they thought would fool other students into thinking they were real sequences.
These students had at least a basic background in probability and were aware of some of the artifacts that were likely in real sequences.

\subsubsection{Initial Deepfakes, the Filtering Generator} 
\cite{Kouritzin} used a method of simulating coin-flip sequences with prescribed pairwise covariances (between flips at two points in the sequence) and marginal probabilities (of heads or tails), which we will call the \emph{simulator}. 
The simulator can, for example, produce a fair sequence where neighbour flips are slightly negatively (positively) correlated and others further away are positively (negatively) correlated to compensate.
Here, the authors reduce the computational requirements of the coin flip simulations by using an explicit formula that encodes the desired covariances. 
Actually, this method turned out to be extendable to discrete (not just binary) random variables on graphs and to be quite useful in applications (see \cite{Kouritzin6}, \cite{Odena}).

We describe the process of simulating deepfakes using the simulator.
Let $r_k$ represent the probability of getting a $1$ on the $k^{th}$ flip $Y_k$ and $\beta^l_{k,j}$ be the covariance $\beta^l_{k,j}=\mathop{\rm cov }(Y_k,Y_{k-j})$.
Then, the model for the trivial faker is:
\begin{equation*}
\text{(TF)}\quad \ r_{k+1}=r_k+\varepsilon r_k(1-r_k)\xi^P_k\quad\text{ and }\quad
\beta^l_{k+1,j}=\beta^l_{k,j}+\varepsilon \beta^l_{k,j}(\beta^l_{k,j}+1)(1-\beta^l_{k,j})\xi^j_k
\end{equation*}
for $k=0,...,199$, $j=1,...,l$.
Here, $\{\xi^P_k\}_{k=1}^\infty$ and $\{\xi^j_k\}_{j,k=1}^{l,\infty}$ are $p=\frac12$, $\{-1,1\}$-Bernoulli independent of everything and $\varepsilon$ is a small parameter.
Fakers that try to undo what has been done follow the random sign change model
\begin{equation*}
\text{(RSC)}\quad \ r_{k+1}=r_k+\rho^P_k(1-2r_k)+\varepsilon r_k(1-r_k)\xi^P_k\quad\text{ and }\quad
\beta^l_{k+1,j}=\rho_{k,j}\beta^l_{k,j}+\varepsilon \beta^l_{k,j}(\beta^l_{k,j}+1)(1-\beta^l_{k,j})\xi^j_k
\end{equation*}
for $k=0,...,199$, $j=1,...,l$.
Here, $\{\rho^P_k\}$ and $\{\rho_{k,j}\}$ are independent such that
$$
P(\rho_{k,j}=-1)=P(\rho^P_k=1)=1-P(\rho_{k,j}=1)=1-P(\rho^P_k=1)=\delta
$$
for some small $\delta>0$.
The trivial faker and the random sign change faker are our initial deepfake models.
Notice both are designed to keep $r_k\in [0,1]$ and $\beta^l_{k,j}\in[-1,1]$
if started that way.
They must be supplied with initial values $r_0\approx \frac12$ (for fairness) and $\beta_{k,j}\approx 0$ for $k-j\le 0$ (for near independence).

The real coin flips can then be modelled in this way and fit into the algorithm
\begin{equation*}
\text{(Real)}\quad \ r_{k}=\frac12\quad\text{ and }\quad
\beta^l_{k,j}=0
\end{equation*}
for $k=1,...,200$, $j=1,...,l$.
Then, for the signal we set
$$
\Theta =\left\{\begin{array}{cc}
    1 & \text{Trivial Faker} \\
    0 & \text{RSC Faker} \\
    -1 & \text{Real Coin}
\end{array}\right.
$$
and let $X_k=\left[\begin{array}{ccc}
   \Theta &
    r_k &
    \{\beta^l_{k,j}\}_{j=1}^l
\end{array}\right]'$
be the signal.
The goal of filtering is to estimate $X_k$ optimally from the back observations
$Y_l,l\le k$, which includes the goal of deepfake detection.
The goal of deepfake detection is to determine the value of $\Theta$ given the observations.

For this method, we initially simulated 137 sequences of 200 coin flips with the above equation and algorithm, which were deemed to be deepfakes due to the known difficulty in distinguishing them from real as well as their computer algorithm generation. 

The algorithm is designed to simulate sequences of coin flips that align with specified marginal probabilities and pairwise covariances. We begin the simulation by defining the total number of flips ($ N_f $) and the number of pairwise covariances ($ N_c $) and the number of pairwise covariances. These parameters dictate the length of the sequence and the complexity of the dependencies between flips. These parameters should be uniformly distributed within specified ranges to ensure variability across trials. Apply the algorithm to simulate a sequence of coin flips. For each flip, compute the probability of it being heads based on the outcomes of the previous flips, then use the uniform random variable $ U $ to determine the flip's result. After generating the sequence, estimate the marginal probability $ \bar{r} $ and the pairwise covariances $ \{\bar{\beta}_j^{Nc}\} $ from the simulated data. These estimates will be used to evaluate how closely the simulated sequence matches the desired statistical properties. Calculate the error for the trial using the formula, $ \text{err} = \frac{(r - \bar{r})^2 + \sum_{j=1}^{Nc} (\beta_j^{Nc} - \bar{\beta}_j^{Nc})^2}{N_c + 1}$. This error metric combines the discrepancies in both the marginal probability and the pairwise covariances, normalized by the number of covariances plus one. By averaging the error over multiple trials, one can assess the algorithm's performance and its ability to generate sequences that accurately reflect the specified probabilities and covariances. The reader is referred to the work \cite{Kouritzin6} for further applied properties and a more detailed explanation of the algorithm.



\subsection{One Step Learning}

After the initial real, fake and deepfake data sequences were set, learning
could begin.

\subsubsection{Generative Learning with GAN}

A Generative Adversarial Network (GAN), a neural-network-based unsupervised learning models developed by \cite{Goodfellow2}, is implemented to simulate deep fake coin flip sequences. 
The GAN model consists of generator and discriminator neural networks (NNs) working simultaneously to create realistic synthetic data in an adversarial set up. 
They have the dual goal to train the generator to produce deepfake sequences that closely mimic real data fooling the discriminator into falsely classifying them as such, as well as to train the discriminator to detect fake sequences as optimally as feasibly possible. 

We design the generator as a feed-forward NN with three dense layers and using Leaky ReLU activations. As an input, a latent vector of 100 independent standard Gaussian (i.e.: $\mathcal N(0, 1)$) random variables is taken, which serves as a noise input. Our generator outputs a binary sequence of length 200, each element of which represents a coin flip. At the output layer, to map the results to binary values, we use sigmoid activation. Throughout the network, batch normalization (at varying levels of momentum) and dropout layers (at varying degrees) are used to stabilize the learning process. We train this network using binary cross-entropy loss and the Adam optimizer. 

For the weights, we manually initialize them for the generator by drawing the weights from a normal distribution with a standard deviation based on the layer dimensions. The intention of this is to avoid issues with randomness associated with random weight initialization, which allows us to avoid sub-optimal convergence and or local maxima. This initialization ensures the generator starts from a more controlled state, which could potentially improve the quality of the deepfakes it produces.

The discriminator is designed as a more complex classification network. It takes a binary sequence of length 200, which may represent a real sequence, handwritten fake sequence, or any of the three deepfake types from the GAN's generator, MOM's generator, or the Bernoulli Generation algorithm . Like the generator, the discriminator consists of dense layers with Leaky ReLU activations, batch normalization, and dropout to prevent overfitting. Instead of a simple binary output, we design the discriminator to classify between the five categories of sequences using a softmax output layer. We train this network using sparse categorical cross-entropy and the Adam optimizer. This balances stability and training efficiency. 

The adversarial tuning of GAN is explained below. The combined initial and adversarial learning of the GAN generator-discriminator is shown in Algorithm \ref{Algorithm:1}. 

\subsubsection{MOM Learning}

As an alternative to GAN, we frame MOM into a GAN-like structure, by replacing both the generator and discriminator neural networks in the traditional GAN framework with MOM components.
The MOM generator is then a pairwise Markov chain with a (potentially
high dimensional) hidden component.
The rates $p$ and $q$, and initial distribution $\mu$ must be learned for each real, fake and deepfake sequence separately as Figure \ref{FIG:1} highlights.

\definecolor{myBlue}{RGB}{246,243,243}

\tikzstyle{block} = [rectangle, draw, fill=myBlue, 
text width=11em, text centered, rounded corners, minimum height=5em]
\tikzstyle{line} = [draw, -latex']

      

\begin{figure}[pos=!htp]
\centering
\resizebox{\textwidth}{!}{%
\begin{tikzpicture}[
    box/.style={draw, rounded corners, thick, align=center, minimum width=5.0cm, minimum height=1.1cm, fill=gray!8},
    bluebox/.style={box, fill=blue!8},
    greenbox/.style={box, fill=green!8},
    orangebox/.style={box, fill=orange!10},
    arrow/.style={->, thick},
    dashedarrow/.style={->, thick, dashed},
    node distance=8mm and 14mm
]

\node[align=center] (h1) at (0,2.4) {\textbf{Training / Learning}};
\node[align=center] (h2) at (8.2,2.4) {\textbf{Generation}};
\node[align=center] (h3) at (16.2,2.4) {\textbf{Classification}};

\node[box] (trainseq) at (0,0.8) {Training sequences by class\$real / handwritten / GAN / simulator, etc.)\\Each sequence has length $N=200$};

\node[bluebox, below=of trainseq] (em) {MOM parameter learning (EM)\\Initialize $(p,q,\mu)$\\Forward/backward recursions + EM updates\\Stop when convergence criterion is met};

\node[box, below=of em] (bank) {Class-specific MOM model bank\\$\{(p_c,q_c,\mu_c): c \in \text{classes}\}$};

\node[greenbox] (gen) at (8.2,-0.1) {MOM sequence generation\\Use fitted $(p,q,\mu)$\\Simulate hidden states and observations\\Output MOM-generated binary sequence};

\node[box, below=of gen] (genseq) {Example generated sequence\\$Y_{\mathrm{MOM}}=(1,0,1,1,0,\dots)$};

\node[orangebox] (testseq) at (16.2,-1.0) {Test sequence (example)\\$Y^\ast=(1,0,1,1,0,\dots)$, length $N=200$};

\node[bluebox, below=of testseq] (score) {Recursive-likelihood / Bayes-factor scoring\\Compute score of $Y^\ast$ under each class model\\in the MOM model bank};

\node[box, below=of score] (label) {Compare scores and assign label\$real / MOM / GAN / handwritten / simulator)};

\draw[arrow] (trainseq) -- (em);
\draw[arrow] (em) -- (bank);

\draw[arrow] (em.east) -- (gen.west);
\draw[arrow] (gen) -- (genseq);

\draw[dashedarrow] (bank.east) -- (score.west);
\draw[arrow] (testseq) -- (score);
\draw[arrow] (score) -- (label);

\end{tikzpicture}
}
\caption{Workflow of the proposed MOM-based framework on a binary-sequence example ($N=200$). The method consists of three stages: (i) MOM parameter learning via EM (forward/backward recursions and updates of $p,q,\mu$), (ii) sequence generation using fitted MOM parameters, and (iii) classification by recursive-likelihood (Bayes-factor) scoring against class-specific model banks, followed by label assignment.}
\label{FIG:1}
\end{figure}

The algorithm for learning the transition probabilities $p$ and $q$ and the initial distribution $\mu$ is given in Algorithm \ref{Algorithm:2} and developed in \cite{kouritzinMarkovObservationModels2025}.
The specifics of the MOM generator are given below.

The discriminator role in MOM is played by recursive likelihood. 
The algorithm to compute recursive likelihood is worked out in
\cite{kouritzinMarkovObservationModels2025}.
The computation and exactly how to use it in the discriminator are stated below.

\subsection{GAN's Generator}
The GAN generator produces deepfake coin flip sequences once fully trained.
It takes takes an input vector consisting of 100 independent standard Gaussian random variable samples, and produces an output deemed a deepfake because it is generated by neither a random number
generator nor coin flip. The GAN generator is first initially trained as a feed-forward NN and then tuned in the adversarial setup (as described below).

The particulars of the GAN generator neural network are as follows: The input layer is fully connected with 1024 neurons and a LeakyReLU activation function with a negative slope coefficient of $\alpha=0.2$. Following this, a dropout layer with a rate of 0.3 is used to prevent overfitting by randomly setting 30\% of the input units to 0 during training. The output is then normalized using a batch normalization layer with a momentum of 0.5 to speed up training. 
Next, the data passes through a second dense layer with 2048 neurons, again with a LeakyReLU activation function ($\alpha = 0.2 $), followed by another dropout layer, again with a rate of 0.3, indicating similar regularization. A second batch normalization layer with a momentum of 0.5 is used to normalize the activations. The output is then flattened into a one-dimensional vector by a flatten layer, preparing it for the final dense output layer. The fully connected output layer has 200 neurons with a sigmoid activation function, ensuring the output values are between 0 and 1, typical for generating binary or normalized data in the range [0, 1]. This neural network is compiled using binary cross-entropy loss and the Adam optimizer with a learning rate of 0.00005.

\subsection{MOM's Generator}

The generator for the MOM solution is a collection of trained pairwise Markov chain models represented by their model parameters $\{p, q, \mu\}$, divided into the
groups real (R), fake (F) and deepfake (D).
Initially, this collection contains only the models obtained from the initial training data.
However, later more models are added from the adversarial setup (as described below).
Actual generation occurs by simulating a \emph{randomly-selected real} MOM model with parameters $p_R, q_R, \mu_R$ say as shown in 
Figure \ref{FIG:2}.

\definecolor{myBlue}{RGB}{246,243,243}

\tikzstyle{block} = [rectangle, draw, fill=myBlue, 
text width=11em, text centered, rounded corners, minimum height=5em]
\tikzstyle{line} = [draw, -latex']

\begin{figure}[pos=!htp]
  \centering
  \resizebox{1\textwidth}{!}{ 
  \begin{tikzpicture}[node distance = 2cm, auto]
    \node [block] (init) {\Large \textcolor{black}{$p_R, q_R, \mu_R $}};
    \node [block, right of = init, node distance = 7cm] (consent) {\large \textcolor{blue}{Simulate MC} $\left(\!\begin{array}{c}X \\Y
    \end{array}\!\right)$};
    \node [block, right of = consent, node distance = 7cm] (refer) {\large \textcolor{black}{Deepfakes (D) $Y$}};
    \path [line] (init) -- (consent);
    \path [line] (consent) -- (refer);
  \end{tikzpicture}
  }
  \caption{Generation in MOM}
  \label{FIG:2}
\end{figure}
The simulation process consists of simulating the pairwise Markov chain and then discarding the hidden component $X$ to produce the generated observations $Y$, which are the deepfake coin flips here.\\
\indent The specifics of the MOM generator are as follows:
The observable component $Y$ is just the sequence of coin flips, represented as ones and zeros.
The hidden layer $X$ is a Markov chain with $s$ states. We start with $s = 6$.
However, during the adversarial process described below it is allowed to increase.
A combined $\begin{pmatrix}
    X & Y
\end{pmatrix}'$ form a Markov chain of dimension $2s$ that has initial distribution 
$\begin{pmatrix}
    X & Y
\end{pmatrix}'\sim\mu$ and one step transition probabilities 
\begin{equation*}P\left(X_t = \hat{x}, Y_t =\hat{y} \mid X_{t-1}=x, Y_{t-1}=y \right)=p_{x\to \hat{x}} q_{y\to \hat{y}}(\hat{x})\end{equation*}
where $p, q, \mu$ have been learned. 

This factorization means that the hidden state process $X_t$ follows a Markov chain with transition probabilities $p_{x\to \hat x}$, while the observation process $Y_t$ depends on both the current hidden state and the previous observation. In particular, $q_{y\to \hat y}(\hat x)$ depends on the previous observation $Y_{t-1}=y$ and the current hidden state $X_t=\hat x$. Thus, unlike a standard HMM, where the observation depends only on the current hidden state, MOM allows the new observation to keep a one-step dependence on the previous observed value \cite{kouritzinMarkovObservationModels2025}.

Like many latent-state models, MOM is not uniquely identifiable in a strict sense, since permuting the hidden-state labels can leave the observed distribution unchanged. This issue can become more apparent when we assume the observation process is binary, where different parameter choices may produce very similar observed sequences. For this reason, we treat MOM mainly as a useful latent model for generation and classification, rather than as a uniquely interpretable description of the underlying process. In practice, we reduce these issues by keeping the number of states small and using multiple initializations.

Note that real coin flips could be simulated in this means by making the initial marginal $\mu_Y$ for $Y$ fair and then using $p, q$ so that each new $Y$ is equally likely to be $1$ or $0$ independent of everything. However, there are more efficient ways.

\section{Methods of Classifying a coin-flip}

This section describes different methodologies of classifying coin flips as real or fake. The different classification methods are: 
human, SVM, GAN discriminator classification, BPF method, and MOM likelihood classification method.

\subsection{GAN Discriminator}

The GAN discriminator, once fully trained, detects (deep)fake coin flip sequences. In particular, the discriminator takes an input vector, consisting of 200 coin flips, and produces an output of one of 5 labels, each corresponding to one of the types of sequences the discriminator is trained on. The discriminator is initially trained as a feed-forward NN and then tuned in the adversarial setup (as described below).

The particulars of the GAN discriminator neural network, optimized to this setting, were determined to be as follows. The input layer is fully connected with 2048 neurons and a LeakyReLU activation function with a negative slope coefficient of 0.2. Following this, a dropout layer with a rate of 0.6 is used to prevent overfitting by randomly setting 60\% of the input units to 0 during training. The output is then normalized using a batch normalization layer with a momentum of 0.5 to stabilize and accelerate training. Next, the data passes through a second dense layer with 1024 neurons, again with a LeakyReLU activation function ($\alpha = 0.2$), followed by another dropout layer, this time with a lower rate of 0.4, indicating less regularization. The fully connected output layer has 5 neurons with a softmax activation function. The model is compiled with a sparse categorical cross-entropy loss function and the Adam optimizer with a learning rate of 0.00005 and a $\beta_1$ set to 0.5 to smooth out the gradient updates by balancing recent gradient values with past values, which is effective for classification problems and deep learning models due to its adaptive learning rate capabilities.

\subsection{Support Vector Machine Classification}

In this section, we investigate the support vector machine (SVM) discriminator in classifying real and fake sequences derived from various sources. SVMs have emerged as a powerful tool for such classification tasks due to their effectiveness in high-dimensional data spaces and capability to model both linear and non-linear decision boundaries. In our study, SVM constructs a decision boundary in a high-dimensional space, which is used to classify sequences into "real" and "fake" classes.
For SVMs, the "best" decision boundary is typically defined as one that maximizes the margin between the classes. This boundary is determined by support vectors—data points from each class that are closest to the decision boundary. For linear classification tasks, the boundary can be understood as a hyperplane that maximizes the margin, while in polynomial classification, this decision boundary can take more complex, non-linear forms in the input space, adapting to the degree of the kernel used.

In SVM, the best hyperplane is defined as one that maximizes the margin between two classes. This distance is defined as the distance between the hyperplane and the support vectors from either class. The support vectors are the closest data points from either class that have the closest distance to the hyperplane. These support vectors define the elements of the hyperplane and model complex decision boundaries. We evaluated SVM's performance based on accuracy, ROC-AUC, AUC-PR, and F1 scores. This study includes an extensive evaluation based on 100 different random seeds, with results consolidated from multiple iterations to ensure robustness. The 100 random seeds were chosen using \texttt{random.randint(1, 100000000)}, which chooses 100 random integers between 1 and 100000000. 

The data set comprises sequences generated from various methods, including simulator sequences, handwritten sequences, MOM sequences, and GAN sequences, which are all highlighted in the "Methods of Data Generation" section.  Additionally, real sequences of length 200 were generated using \texttt{random.randint(0,1)} from Python's \texttt{random} package, which generates each element in a sequence as a random binary value (0 or 1). We processed each set of sequences into a combined dataset with labels "fake" (0) and "real" (1).

We used polynomial kernels for the classification tasks. The model was trained on the training set subsetted from splitting the full data into an 80-20 split. We evaluated on the testing set using the accuracy, ROC-AUC, AUC-PR, and F1 score metrics. The performance was assessed across varying degrees of polynomial kernels (1, 2, 3, 5, 7, 10, 15) and various random seeds to ensure robustness of the results. 

\subsection{Filtering Classification}

The filtering approach as it stands did not require learning. The reason for this is it uses specific real and faker mathematical models. It is true that the faker model has static parameters that could be learnt
but these were learnt in an earlier work and regardless, are just numbers
that can be learnt one time offline.
The generator for the filtering approach is just a random number generator and to include deepfakes the Simulator algorithm given above.

We represent the attributes of the faker and the real coin in the mathematical models used in the Simulator algorithm and filtering approach. 
We do this through pairwise covariance and marginal probabilities between each flip and the flips that came before it in time \cite{Kouritzin}. 
In filtering terms, the models of the real coin and the fakers with flips have termed signals, and the sequence of coin flips is called observations. 
We use a filtering technique to generate real-time estimations of the likelihood of various competing signal models based on the observations. We present the problem in a particle filtering framework (see Algorithm \ref{Algorithm:3}), describe how we obtained faked data and our observations of its properties, discuss algorithms for simulating flip sequences from marginal probabilities and pairwise covariances, and present empirical results of our implementation of the filtering solution. The fake coin identification problem is presented here within the framework of a filtering algorithm. We need accurate and effective models of both the observations and the possible signals. The signals in this problem are time-inhomogeneous marginal probabilities and covariances, along with a real coin or faker-type indication. We employed the combined branching approach, which is explained in the paper \cite{Kouritzin1}. Classifying, tracking, and predicting the signal based on observations is the aim of filtering. The Branching Sequential Monte Carlo algorithm was described to reduce particle number fluctuations and thereby improve performance and reliability. We have used three signal $\{X^i_t, t = 1, 2,\dots\}^3_{i=1}$ with the associated weights $L^i_t$ , as mentioned in the paper \cite{Kouritzin}. $\sigma^N$ is an approximation of the unnormalized filter in problems like tracking and model selection which measured in such way $\sigma_t^N = \frac{1}{N} \sum\limits_{j = 1}^{N_{t-1}} \hat{L}^j_t \delta_{\hat{X}^j_t}$ where $\delta$ denotes dirac measure and $\hat{X}^j_t$ denotes the path of the $j^{th}$-particle In the branching algorithm, when the prior weight $\hat{L}^j_t$ for particle $j$ is outside of a certain interval around the average weight, we do the branching, which helps to preserve the process distribution. 
In particle filtering, the resampling parameter $r$ and the average particle weight $A_t$  are crucial for maintaining effective particle diversity and avoiding degeneracy, where too few particles carry meaningful weights. The parameter $r$ defines a threshold range around $A_t$, which is the mean of all particle weights at each time step. This range helps to determine whether particles with significant weight deviations should be resampled. Resampling replicates higher-weight particles while discarding lower-weight ones. When weights are close to $A_t$, resampling may not be needed, but larger deviations indicate that resampling can help reduce the impact of weight disparity. Adjusting $r$ based on weight variance can further refine the balance between computational efficiency and filter accuracy \cite{Doucet,Cappe}. In our study, with $r$ set to $4.5$, the particle filter uses this specific threshold to determine which particles should be resampled based on how far their weights deviate from the average weight $A_t$. This means that particles with weights differing significantly (beyond 4.5 times the average weight) are targeted for resampling, which helps maintain a balanced distribution of particle weights and improves the filter's accuracy.

Particles that are branched result in zero or more particles, which are assigned the average weight $A_t$, are added at the same location as the parent. In other words, we copy the path with extreme prior weight and give the copies, if there are any, the current average weight. When its prior weight $\hat{L}^j_t$ is not extreme, a particle is not branched and gets to keep its prior weight. 
The resampling parameter $r$ determines the size of the interval around the average weight $A_t$  outside of which particles are considered extreme and are branched. The distance between the estimates of  value of marginal probabilities and pairwise covariance, and the real coin is used as an error metric; a threshold on the error metric is then empirically determined \cite{Kouritzin}. For a given sequence, if the error metric falls within the threshold, then the sequence is real; otherwise it is fake.

\subsection{MOM's Discriminator}
The likelihood is a crucial concept in Bayesian statistics, used to quantify the evidence for one model against another. The recursive likelihood from \cite{kouritzinMarkovObservationModels2025} provides a way to update the odds for competing hypotheses based on observed data. 
The recursive likelihood's ability to incorporate prior information and provide a continuous measure of evidence makes it a versatile and powerful tool for hypothesis testing. One significant advantage of using this is for the interpretability. However, calculating the likelihood can be challenging, especially for complex models with high-dimensional parameter spaces. Advances in computational techniques, such as Markov Chain Monte Carlo (MCMC) methods and variational inference, have made it more feasible to compute likelihoods for a wider range of models.

\definecolor{myBlue}{RGB}{246,243,243}

\tikzstyle{block} = [rectangle, draw, fill=myBlue, 
text width=11em, text centered, rounded corners, minimum height=5em]
\tikzstyle{line} = [draw, -latex']

\begin{figure}[pos=!htp]
  \centering
  \resizebox{1\textwidth}{!}{ 
  \begin{tikzpicture}[node distance = 2cm, auto]
    \node [block] (init) {\Large \textcolor{black}{Unknown Sequence}};
    \node [block, right of = init, node distance = 7cm] (consent) {\large \textcolor{blue}{Likelihood}};
    \node [block, right of = consent, node distance = 7cm] (refer) {\large \textcolor{black}{Classified Sequence}};
    \node [block, below of = consent, node distance = 2.7cm] (refer1) {\large \textcolor{black}{$p_R, q_R, \mu_R $\\$p_F, q_F, \mu_F $\\$p_D, q_D, \mu_D $}};
    \path [line] (init) -- (consent);
    \path [line] (refer1) -- (consent);
    \path [line] (consent) -- (refer);
  \end{tikzpicture}
  }
  \caption{Discriminator in MOM model}
  \label{FIG:3}
\end{figure}

The MOM's discriminator utilizes the recursive likelihood to distinguish the type of sequence being inputted. For each sequence inputted into the discriminator, we compute a recursive likelihood between the sequence and every $p, q, \mu$ model generated earlier. Each recursive likelihood computed is assigned a label, depending on the type of sequence the $p, q, \mu$ model used is from. To classify the sequence, we split the labels and corresponding recursive likelihoods into 2 categories: those that came from a $p, q, \mu$ model generated from a sequence of the same type as the input, and the other being the labels that came from the other models. We then take the average of the recursive likelihoods in each group, and classification of the sequence is "correctly identified" or "incorrectly identified", depending on which average is greater. 

To integrate the recursive likelihood as a discriminator, we need to follow a process where we calculate the recursive likelihood for each input sequence against pre-trained models and use these factors to classify the sequence as its certain type.  Train multiple $( p, q, \mu )$ models using the observed sequences.
For each input sequence, calculate the recursive likelihood using the pre-trained models. Assume we have trained models Real, GAN, SVM, simulator, MOM. 
For an input sequence $Y$ and each $(p, q, \mu)$ model from the generation portion, compute the recursive likelihoods of $Y$ with each model. Each recursive likelihood value computed is assigned a label based on the type of sequence the model is known to generate ("Real" for real data sequences, "MOM" for MOM deepfakes, "GAN" for GAN deepfakes, "simulator" for simulator fakes, and "Handfakes" for handwritten fakes). The final classification of the sequence is determined by the maximum between the average of the recursive likelihoods with the correct label, and average of the recursive likelihoods with the incorrect labels. Should the former be greater, the sequence is classified as "correctly identified", and "incorrectly identified" otherwise. 

By using the recursive likelihood in the MOM discriminator, we leverage the statistical evidence from multiple models to classify sequences. The process involves training multiple models, computing the recursive likelihoods, assigning labels, and using the top recursive likelihood labels to make a final classification decision. This method provides a robust mechanism to distinguish between real and fake sequences based on the collective evidence from several models.
\section{Adversarial Tuning}

As suggested in the name, GAN is an adversarial process which combines the functions of the generator and the discriminator. The GAN takes a latent noise vector as input through the generator, which then generates synthetic sequences. These sequences are passed through the discriminator to classify them as real or fake. Our GAN model is compiled with binary cross-entropy loss and the Adam optimizer. We pre-set the maximum number of epochs to be 500 and the batch size to 64. Within each epoch, the system generates both real and fake samples, trains the discriminator based on these generated samples and the deepfakes, and simultaneously trains both the generator and the discriminator based on their respective loss functions.
\\
For the discriminator, sparse categorical cross-entropy (SCE) loss is used to measure how well the network can classify sequences into one of five categories: real (0), GAN-generated (1), MOM-generated (2), handwritten (3), and simulator (4) sequences. Let $N$ be the number of training samples, $x_i$ the input sequence, $y_i \in \{0, 1, 2, 3, 4\}$ be the true label of the sequence, and $p_\theta(y_i \mid x_i)$ be the discriminator's predicted probability for the correct class, where $\theta$ is the current parameter values of the discriminator. The SCE loss is then defined as 
\begin{equation}\label{sce}
    \mathcal L_D = -\frac{1}{N} \sum_{i=1}^N \log p_\theta(y_i \mid x_i)
\end{equation}
Given the true labels and the discriminator's predictions, this loss function quantifies the difference between the actual and predicted labels across these five classes. Minimizing $\mathcal L_D$ improves the discriminator's ability to correctly classify the type of sequence it receives. By adjusting the network's weights during training, the discriminator learns to more accurately distinguish between real sequences, handwritten sequences, and the various deepfakes.
\\
For the generator, binary cross-entropy (BCE) loss is used. Let $z_i$ be the random latent noise input, $G(z_i)$ be the generator output (i.e. GAN-generated), $D(G(z_i))$ be the discriminator's probability that $G(z_i)$ is real. The BCE is 
\begin{equation}\label{bce}
    \mathcal L_G = -\frac{1}{N}\sum_{i=1}^N \left[y_i \log(D(G(z_i))) + (1 -y_i) \log(1 - D(G(z_i)))\right]
\end{equation}
Here, the generator's goal is to produce realistic deepfake sequences to fool the discriminator into outputting a high probability of them being real, close to 1. Therefore, the BCE loss of the generator is calculated with "opposite" labels, where it aims to minimize the loss between its fake output (which it wants the discriminator to classify as real) and the real label.
\\
Minimizing each network's respective loss functions essentially leads to a more effective adversarial relationship between the generator and the discriminator. This drives the GAN to improve both the generator's ability to synthesize realistic data and the discriminator's ability to correctly classify the data as real or fake. As per \cite{Goodfellow}, GAN training is usually expressed as a min-max game: 
\begin{equation*}
    \min_G\max_D V(D, G) = \mathbb E_{x \sim p_{\text{data}}}[\log(D(x))] + \mathbb E_{z \sim p_z}[\log(1 - D(G(z)))]
\end{equation*}

The model is compiled with sparse categorical cross-entropy (SCE) loss for the discriminator and binary cross-entropy (BCE) loss for the generator, both of which are optimized using the Adam optimizer. The use of SCE allows the discriminator to classify sequences into one of five categories, while BCE ensures that the generator’s goal is to produce deepfake data that is realistic enough to fool the discriminator into classifying the fake data as real. During training, the discriminator should ideally output a high probability for real data and a low probability for the fake sequences generated by the GAN.

\section{Comparative Results}

In this section, the table \ref{tbl1} presents the study how the real, handwritten, and sequences generated from a Simulator, GAN and MOM are classified using the GAN discriminator, support vector machines (SVM), particle filtering, and the MOM recursive likelihood approach. In other words, we want to study whether MOM performs better in classifying different types of sequences than the other methods. All methods perform multiclass classification over the five sequence types: Real, Handwritten, Simulator, GAN, and MOM. For each method, we perform studies for the handwritten fake sequences,  and the other generated sequences. There are 137 fake sequences of each type, each of length 200. The number of real sequences we generate corresponds to however many fake sequences we consider (either 137 or 548).

Across different polynomial degrees, the SVM model performs best with lower-degree kernels (1, 2, and 3) for generating sequences like GAN and Handwritten Sequences, with some variation in performance on more complex types like Tricky and MOM sequences. The overall accuracy improves up to Degree 2 but fluctuates afterwards, with Degree 5 showing some promise. However, higher degrees (7 and 10) give worsening results, suggesting overfitting and increased complexity that the model cannot handle effectively. Standard deviations are consistently low, indicating that the model performs consistently across runs, but it struggles to generalize across all sequence types. The results suggest that while SVM can be tuned to achieve higher accuracy on specific types of sequences, its overall performance across different types of fakes is relatively stable but not particularly high. 

For GAN's training, we split the types of sequences into training and testing sets using an 80:20 split. We train on 2000 epochs using a batch size of 64. For testing, we feed each sequence in the testing set into the trained discriminator. In table \ref{tbl1}, the GAN shows mixed results. It performs poorly on a Bernoulli Generation algorithm (6.18\%) and MOM (7.36\%) sequences, indicating difficulty in accurately detecting these types. However, it achieves moderate success on GAN sequences (54.96\%) and performs reasonably well on Handwritten Sequences (74.64\%) and Real sequences (74.33\%). Overall, the GAN model achieves 54.88\% accuracy, but the relatively high standard deviations (ranging from 2.87 to 9.41) suggest inconsistent performance, making the model less reliable across different sequence types . This indicates that while the GAN discriminator is moderately effective at distinguishing real sequences, its ability to detect fake sequences, particularly handwritten ones, is less robust, showing significant variability in its performance.

The particle filtering applied to different coin flip sequences, specifically handwritten and all generated (MOM, GAN,  Bernoulli Generation algorithm \& handwritten) sequences. The average accuracy is used to evaluate the performance of the particle filtering algorithm. The Particle Filtering demonstrates strong and consistent performance across all sequence types. With accuracies ranging from 80.67\% (Handwritten Sequences) to 90.88\% (GAN sequences), it shows proficiency in handling a variety of sequence types. The overall accuracy is 86.32\%, making it one of the top-performing models. Importantly, the model has very low standard deviations (between 0.08 and 0.09 for each type), indicating highly stable and reliable performance. Overall, these results highlight the efficacy of the particle filtering algorithm in accurately distinguishing between different types of coin flip sequences, maintaining high accuracy and low variability across both handwritten and other fake sequences. This consistency in performance underscores the algorithm's potential reliability and effectiveness in practical applications involving sequence analysis and classification. 

For MOM, similar to the GAN, we split the sequences into training and testing sets using an 80:20 split. We first consider a hidden layer consisting of $s$ states.  To initiate our model, we consider a "canonical model" and consider a case where the hidden layer has 2 states and a $q$ matrix that mimics a "perfect" coin flip (where all entries are $0.25$). Consider a randomly generated $p$ matrix and a $\mu$ matrix with 0.25 as every entry. Using these matrices, we generate the canonical model and use this to initialize the set of real sequences. The rest of the real sequences are randomly generated using Python's \texttt{random} package. Using the EM algorithm outlined in \cite{kouritzinMarkovObservationModels2025}, the $p, q, \mu$ transition matrices of the real sequences generated by Python's \texttt{random} module are maximized starting from $6$ hidden states. Then, a subset of the maximized $p, q, \mu$ matrices of the real sequences are used to simulate more coin flip sequences, which we refer to as "generated deepfakes" data. Post-generation, we increase the hidden states to $7$ and re-generate new  $p, q, \mu$ models of the real sequences, while generating the $p, \mu$ matrices of the generated and fake sequences following the same procedure as outlined above. The algorithm \ref{Algorithm:2} describes how to produce real, generated, and fake sequences. Note that fake sequences are created by hand to mimic coin flips as realistic as possible. Here, we present an algorithm to generate and analyze sequences of coin flips as realistically as possible. 

For the training portion of the MOM, we initially consider a hidden layer consisting of six states. After 100 trials running against sequences, the MOM exhibits the highest overall accuracy, with an impressive 92.27\%. It performs exceptionally well on a Bernoulli Generation algorithm  (99.82\%), GAN (97.25\%), and Real (98.76\%) sequences, showing that it is highly effective in detecting most types of sequences. However, its accuracy on MOM sequences is lower (76.59\%), suggesting some difficulty in detecting sequences of its own type which shows how strong the MOM generation is. While the model generally performs well, the higher standard deviation for MOM sequences (13.53) indicates variability in its performance for this category.  These results indicate that MOM is not only highly effective at distinguishing between real and fake sequences, but also particularly strong at deepfake detection, especially compared to GAN. Table \ref{tbl1} provides the results of the performances of each classification method against each type of sequence, as well as the overall accuracy of each method. 

Complete implementation details for GAN, MOM, and BPF (including hyperparameters, initialization, stopping criteria, and runtime settings used for Table \ref{tbl1}) are provided in Appendix \ref{app:implementation}.

\subsection{Experimental Protocol}\label{sec:exp-protocol}
Each sequence has length $T=200$. We consider five sequence types: real, handwritten fake, and fake sequences generated by the Simulator, GAN, and MOM. For each fake type, we construct a balanced binary classification dataset by matching the number of real sequences 1:1 to the number of fake sequences (137 real and 137 fake sequences). For pooled experiments across all four fake types, we use 548 real sequences to match the 548 fake sequences.

Unless otherwise stated, all classifiers (SVM, GAN discriminator, particle filtering, and MOM discriminator) use the same stratified 80:20 train/test split at the sequence level. For each repeated run, the train/test split is generated independently, and all methods are evaluated on the same split for fairness. Hyperparameters are selected using training data only (via a validation split within the training set), and the test set is used only for final evaluation.

We report mean accuracy and standard deviation over 100 repeated runs with different random seeds.

\begin{table}[
pos = !htp]
\caption{Classification performance by sequence type in the 5-class setting. The sequence types are: \emph{Real}, computer-generated fair coin-flip sequences treated as authentic; \emph{Handwritten Seq.}, human-produced fake coin-flip sequences; \emph{Simulator}, deepfake sequences produced by the Bernoulli-generation simulator; \emph{GAN}, deepfake sequences generated by the GAN generator; and \emph{MOM}, deepfake sequences generated by the Markov Observation Model. For each method, the entries in the five sequence-type columns report the held-out classification accuracy for sequences whose true label is the indicated class. The \emph{Overall} column reports pooled accuracy on the combined held-out evaluation set across all five sequence types, not a macro-average.}
\label{tbl1}
\begin{center}
    Sequence Type
\end{center}
\begin{tabular*}{\tblwidth}{@{} LCCCCCC@{} }
\toprule 
 & \textbf{Simulator($\%$)} & \textbf{GAN($\%$)}    & \textbf{Handwritten Seq.($\%$)}  & \textbf{MOM($\%$)} & \textbf{Real($\%$)} &  \textbf{Overall($\%$)} \\
\midrule
\textbf{SVM} & & & & &\\
\textbf{Degree 1}\\
Accuracy  & 53.31 & 73.89 & 60.98 & \textbf{\textcolor{orange}{50.1}} & 51.09 & 47.42\\
\scriptsize Standard deviation  & \scriptsize 0.05 & \scriptsize 0.02 & \scriptsize 0.05 & \scriptsize 0.03 & \scriptsize 0.08 & \scriptsize 0.06\\
\\
\textbf{Degree 2}\\
Accuracy  & 53.32  & 71.33 & 69.33  & \textcolor{orange}{\textbf{50.4}} & 58.81 & 52.7 \\
\scriptsize Standard deviation  &  \scriptsize 0.05 & \scriptsize 0.02 & \scriptsize 0.08 & \scriptsize 0.03 & \scriptsize 0.05 & \scriptsize 0.01 \\
\\
\textbf{Degree 3}\\
Accuracy  & 47.43 & 75 & 70.06 & \textcolor{orange}{\textbf{51.5}} & 50.96 & 49.34 \\
\scriptsize Standard deviation  &  \scriptsize 0.06 & \scriptsize 0.04 & \scriptsize 0.06 & \scriptsize 0.01 & \scriptsize 0.01 & \scriptsize 0.01\\
\\
\textbf{Degree 5}\\
Accuracy  & 51.82 & 74.27 & 71.19 & \textcolor{orange}{\textbf{48.6}} & 49.2 & 50.88 \\ 
\scriptsize Standard deviation  &  \scriptsize 0.03 & \scriptsize 0.07 & \scriptsize 0.05 & \scriptsize 0.04 & \scriptsize 0.02 & \scriptsize 0.01\\
\\
\textbf{Degree 7}\\
Accuracy  & 52.59 & 74.63 & 52.59 & \textcolor{orange}{\textbf{48.9}} & 49.67 & 50.83\\
\scriptsize Standard deviation  &  \scriptsize 0.04 & \scriptsize 0.04 & \scriptsize 0.03 & \scriptsize 0.03 & \scriptsize 0.02 & \scriptsize 0.01\\
\\
\textbf{Degree 10}\\
Accuracy  & 49.26 & 53.29 & 51.09 & \textcolor{orange}{\textbf{47.1}} & 49.7 & 51.82 \\
\scriptsize Standard deviation  &  \scriptsize 0.01 & \scriptsize 0.02 & \scriptsize 0.01 & \scriptsize 0.05 & \scriptsize 0.02 & \scriptsize 0.01\\
\\
\midrule
\textbf{GAN}\\
Accuracy & 6.18 &54.96 &74.64 & \textcolor{orange}{\textbf{7.36}}& 74.33 & 54.88\\
\scriptsize Standard deviation  & \scriptsize 4.89 & \scriptsize 9.41 & \scriptsize 5.91 & \scriptsize 4.42 & \scriptsize 5.11 & \scriptsize 2.87\\

\midrule
\textbf{Particle filtering} & & & & &\\
Accuracy  &86.86  &80.67  &\textcolor{orange}{\textbf{90.88}} & \textcolor{orange}{\textbf{73.67}} &87.96 &86.32\\
\scriptsize Standard deviation  &\scriptsize 0.025  &\scriptsize 0.3  &\scriptsize 0.012 &\scriptsize 0.024 &\scriptsize 0.018 &\scriptsize 0.010\\
\midrule
\textbf{MOM} & & & & &\\
Accuracy  & \textcolor{orange}{\textbf{99.82}} & \textcolor{orange}{\textbf{97.25}} & 89.29 & \textcolor{orange}{\textbf{76.59}} & \textcolor{orange}{\textbf{98.76}} & \textcolor{orange}{\textbf{92.27}}\\
\scriptsize Standard deviation  & \scriptsize 0.78 & \scriptsize 3.89 & \scriptsize 2.43e-13 & \scriptsize 13.53 & \scriptsize 2.05 & \scriptsize 2.93\\
\bottomrule
\end{tabular*}
\end{table}

\subsection{Human Baseline Mini-Study}
To provide a baseline for comparison with machine learning methods, we conducted a small study observing human performance on the coin flip discrimination task. A total of 83 respondents were presented with 30 binary sequences: 10 \textit{real}, 10 \textit{handwritten}, and 10 \textit{MOM generated deepfakes}. For each sequence, respondents were asked to classify it as real, handwritten, or deepfake.

\begin{figure}[pos=!htp]
    \centering
    \includegraphics[scale=0.1]{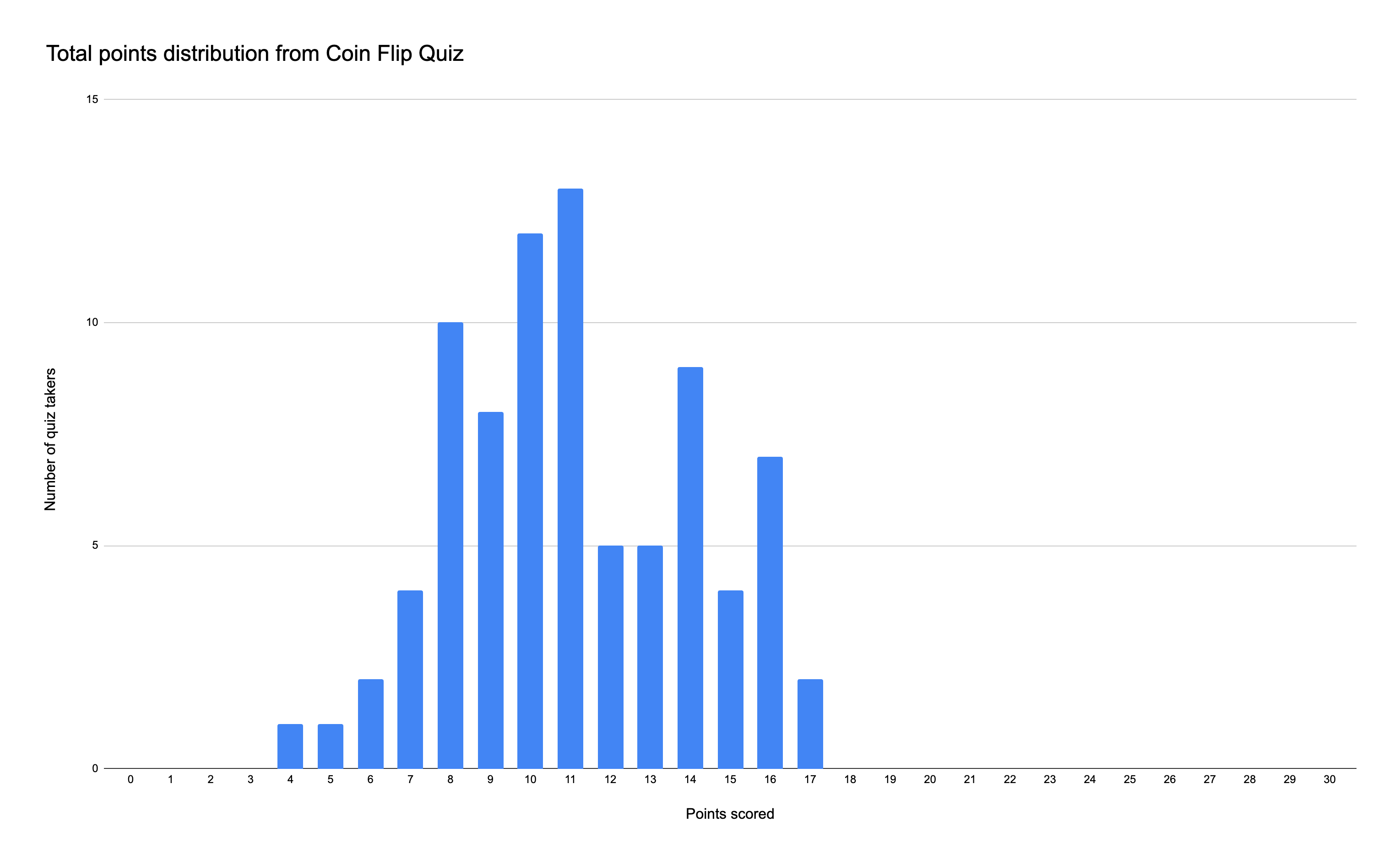}
    \caption{Distribution of the results of the coin flip quiz}
    \label{fig:quizhist}
\end{figure}

Figure \ref{fig:quizhist} shows the distribution of total quiz scores. The mean score was 11/30 (median = 11), only slightly above chance performance.  The majority of participants only scored between 10 and 17, with no score being higher than 17, and the lowest score being 4. 

The aggregate classification results are summarized in the confusion matrix Table \ref{tab:confusionmatrixquiz}. Each row corresponds to the true sequence type, and each column corresponds to the human classification. The rows each sum to 830, which reflects the 10 sequences of each type evaluated by the 83 respondents. 

\begin{table}[
pos = !htp]
    \caption{Confusion matrix of the aggregate quiz results}
    \label{tab:confusionmatrixquiz}
    \centering
    \begin{tabular}{c|ccc}
        \textbf{True/predicted} & \textbf{Predicted Real} & \textbf{Predicted Handwritten} & \textbf{Predicted Deepfake}  \\
        \hline
        \textbf{True Real} & 320 & 260 & 250 \\
        \textbf{True Handwritten} & 236 & 333 & 261\\
        \textbf{True Deepfake} & 310 & 252 & 268 
    \end{tabular}
\end{table}

Notice the emergence of several patterns. Respondents were most successful at identifying human-written fakes (around 43.3\% accuracy), which suggests that such sequences contain detectable artifacts to the human eye. In contrast, performance was much poorer on distinguishing real from deepfake sequences: only around 40.2\% of real sequences were correctly identified, with nearly as many (around 31.4\%) misclassified as deepfakes. Similarly, only around 35.2\% of deepfakes were correctly labeled, with many misclassified as real (38.1\%).

In the end, participants did perform above chance but far from reliably so. Their relative success at spotting handwritten fakes was contrasted with their near chance accuracy in distinguishing real sequences from deepfakes. Here, we draw the conclusion that the human eye is poorly suited for detecting algorithmically generated randomness, which emphasizes the value of the existence of probabilistic and machine learning methods for deepfake detection.

\section{Conclusions and Future Work}
Based on our comparative analysis, several key insights emerge regarding the creation and detection of deepfakes using different classification methods. The GAN discriminator, while moderately effective, demonstrates significant variability in its ability to detect fake sequences, particularly handwritten ones. Its overall accuracy drops considerably when tested against all types of fake sequences, indicating limitations in generalizability and robustness. This suggests that while GANs can generate realistic sequences, their discriminators require further refinement to reliably distinguish between real and fake data. The SVM's performance improves with the degree of the polynomial to a point, achieving the highest accuracy with a degree of 5 for handwritten sequences. However, its performance declines sharply with higher degrees and remains relatively stable but not particularly high when tested against all types of fake sequences. This indicates that while SVMs can be fine-tuned for specific types of deepfake detection, their overall efficacy across diverse fake sequences is limited. The particle filtering method shows impressive accuracy and consistency in classifying both handwritten and all types of fake sequences. Its high accuracy and low variability underscore its potential reliability and effectiveness in practical applications involving sequence analysis and classification. This method demonstrates robustness in detecting deepfakes generated by various methods, making it a valuable tool in the fight against deepfake fraud. MOM outperforms the other methods significantly, achieving near-perfect accuracy for both real and handwritten sequences and maintaining high performance against all types of fake sequences. Its minimal performance variability and consistently high accuracy highlight its robustness and efficacy in distinguishing between real and fake sequences. 

In addition to accuracy, the methods differ substantially in computational cost. The asymptotic complexity analysis in Appendix \ref{app:complexitycomparison} shows that the BPF scales linearly in sequence length and particle count, while the GAN has a higher up-front training cost but relatively cheap classification cost once the discriminator is trained. By contrast, MOM is computationally heavier than both because it combines repeated EM fitting, multiple random initializations, and model-bank scoring at test time. In our implementation, this additional computational cost was accompanied by stronger classification performance, indicating a practical trade-off between efficiency and accuracy in the controlled setting considered here. 

In summary, while GAN offers some utility in deepfake detection, its performance is less reliable and variable. Particle filtering shows robust performance, and MOM stands out as the most effective method, demonstrating high accuracy and consistency. Future work should focus on enhancing the robustness and generalizability of GANs and SVMs, exploring the integration of particle filtering and MOM techniques, and developing new methods to improve deepfake detection accuracy further. Additionally, research should consider the evolving complexity of deepfake generation techniques to ensure detection methods remain effective against increasingly sophisticated fakes. Further exploration into hybrid models combining strengths from multiple techniques could also provide more comprehensive solutions for deepfake detection.

\bibliographystyle{plain}  
\bibliography{coinflip}

\section*{Appendices}

\appendix
\section{Implementation Details and Reproducibility}\label{app:implementation}
\subsection{GAN Implementation Details}\label{app:gan-details}
 The GAN discriminator is trained as a 5-class classifier with labels
\begin{equation*}
0=\text{real},\quad 1=\text{GAN},\quad 2=\text{MOM},\quad 3=\text{handwritten},\quad 4=\text{simulator}
\end{equation*}
Let $x_i$ denote an input sequence and $y_i\in\{0,1,2,3,4\}$ its class label. The discriminator parameters $\theta_D$ are trained using sparse categorical cross-entropy:
\begin{equation*}
\mathcal{L}_D(\theta_D)
=
-\frac{1}{N}\sum_{i=1}^N \log p_{\theta_D}(y_i\mid x_i).
\end{equation*}
In each discriminator update, the loss is computed separately for real, GAN-generated, handwritten, simulator, and MOM-generated mini-batches, and the five losses are averaged.

The generator is trained through the combined GAN model using sparse categorical cross-entropy with target label $0$ (the ``real'' class) (i.e., it is optimized to generate sequences that the discriminator classifies as real).

The generator takes a latent vector of dimension 100 and outputs a binary sequence of length $200$ (via a sigmoid output layer followed by thresholding at 0.5 for generation). The generator uses two hidden dense layers (1024 and 2048 units) with LeakyReLU activations, dropout (rate 0.3), and batch normalization (momentum 0.5). The discriminator uses dense layers of sizes 2048 and 1024 with LeakyReLU activations, dropout (rate 0.3), and batch normalization (momentum 0.5), followed by a 5-class softmax output.

For the final experiments reported in Table \ref{tbl1}, we used:
\begin{itemize}
    \item Adam optimizer for both generator and discriminator,
    \item learning rate $10^{-5}$,
    \item $\beta_1=0.5$,
    \item batch size 64,
    \item 500 epochs,
    \item one discriminator update and two generator updates per epoch.
\end{itemize}
No learning-rate schedule was used.

To improve training stability, we used dropout and batch normalization in both networks, and added Gaussian noise with standard deviation 0.1 to real and generated sequences during discriminator training. We did not use additional anti-collapse penalties (e.g., gradient penalty or feature matching).

The GAN generator outputs a length $200$ vector with sigmoid activations. To obtain a binary sequence, each coordinate is thresholded at $0.5$: values greater than $0.5$ are mapped to $1$, and values less than or equal to $0.5$ are mapped to $0$. No temperature scaling, Gumbel-softmax, or other stochastic sampling rule is used at generation time. The latent input is sampled i.i.d. from a 100-dimensional standard Gaussian distribution.

During discriminator training, additive Gaussian noise with standard deviation $0.1$ is applied to both real and GAN-generated sequences before the discriminator update. This same noise setting is used in all runs. We did not use a separate temperature-based or stochastic label-smoothing schedule. 

\subsection{MOM Implementation Details}\label{app:mom-details}
The MOM experiments were run over 100 random seeds. For each run, the hidden-state dimension was initialized at $s=6$ and the observation was set to be binary. The state-size schedule was fixed (not adaptively selected): we first fit models with $s=6$ to generate MOM-based fake sequences, then increased the hidden-state dimension once to $s=7$ and refit class-specific models used for classification.

In the MOM implementation, real sequences were modelled with length $N_{\text{real}}=400$, while fake sequences (MOM-generated, handwritten, tricky, and GAN-generated) were modelled with length $N_{\text{fake}}=200$. We used an 80:20 train/test split for each class.

For each fitted model:
\begin{itemize}
    \item $P$ was initialized as a random row-stochastic matrix by sampling each row from i.i.d.\ Uniform$(0,1)$ values and normalizing
    \item $Q$ was initialized from empirical transition counts of the observed sequence, followed by random perturbation of nonzero entries and row normalization
    \item $\mu$ was initialized randomly, with entries incompatible with the first observation forced to zero, followed by normalization to sum to one
\end{itemize}

Parameter updates were performed using forward-backward recursions and iterative updates of $(P,Q,\mu)$. For each fit, we used a maximum of 1000 iterations. The stopping criterion was based on parameter changes (not log-likelihood): we stopped when
\begin{equation*}
\sum_{i,j} |P^{(m)}_{ij}-P^{(m-1)}_{ij}| < 5\times 10^{-3}
\quad\text{and}\quad
\sum_{i,y} |\mu^{(m)}_{iy}-\mu^{(m-1)}_{iy}| < 5\times 10^{-3}.
\end{equation*}
For each training sequence and each class, we generated 10 random initializations (denoted \texttt{model1}-\texttt{model10}) and fit the corresponding MOM parameters. For each fitted model, we computed a recursive likelihood. In the implementation used for the reported experiments, the score was recorded for all 10 fits and the final fitted model in the loop was retained for later usage. 

For each test sequence, we evaluated Bayes-factor scores against all fitted model banks (real, MOM-generated, handwritten, ricky, and GAN). Class-specific decisions were made by comparing the mean recursive likelihood over models from the target class against the mean recursive likelihood over models from the remaining classes.

\subsection{Branching Particle Filter (BPF) Implementation Details} \label{app:bpf-details}

For the branching particle filter, we initialized the filter with 200 particles and averaged the reported results over 100 independent runs to reduce Monte Carlo variability.

Branching was controlled using a threshold parameter $r$ in the range $3 \le r \le 3.5$ (as used in the interval test in Algorithm \ref{Algorithm:3}). In our experiments, this range provided stable performance.

We also tested larger particle counts and observed little improvement in accuracy relative to 200 particles, while computational cost increased noticeably. For this reason, we used 200 particles in the final reported experiments.

\section{Runtime and Computational Cost}\label{app:complexitycomparison}
For the GAN (shown in Algorithm \ref{Algorithm:1}), let 
\begin{itemize}
    \item $E$ be the number of epochs
    \item $M$ be the number of mini-batches per epoch
    \item $B$ be the batch size
    \item $N$ be the sequence length ($N= 200$)
    \item $d_z$ be the latent dimension ($d_z = 100$)
    \item $C$ be the number of classes in the discriminator ($C = 5$)
\end{itemize}
Let $W_G$ and $W_D$ denote the per-sample dense-layer operation counts for generator and discriminator. In our architecture,
\begin{align*}
W_G &= O(d_z\cdot 1024 + 1024\cdot 2048 + 2048\cdot N)\\
W_D &= O(N\cdot 2048 + 2048\cdot 1024 + 1024\cdot C)
\end{align*}
Each discriminator update processes five class-specific batches (real, GAN, MOM, handwritten, simulator), giving $O(BW_D)$ work per mini-batch up to a constant factor. A generator update through the combined GAN model costs $O(B(W_G+W_D))$. Therefore, the total GAN training complexity is
\begin{equation*}
O(E(MBW_D + B(W_G+W_D)))
\end{equation*}

For the MOM (shown in Algorithm \ref{Algorithm:2}), let 
\begin{itemize}
    \item $N$ be the sequence length
    \item $s$ be the hidden-state dimension
    \item $K$ be the number of EM iterations (until stopping, capped by the stopping limit)
    \item $R$ be the number of random initializations per sequence (in our implementation, $R=10$)
    \item $J_1$ be the number of sequences fit in the initial state-$s$ stage
    \item $J_2$ be the number of sequences fit after raising the state dimension to $s+1$
    \item $L$ be the number of stored MOM models used for recursive-likelihood (Bayes-factor) scoring at test time
\end{itemize}

In one EM iteration, the dominant cost comes from the forward and backward recursions and the parameter updates for $p,q,\mu$. Each of these scales as $O(Ns^2)$ . Therefore, one EM iteration costs
\begin{equation*}
O(Ns^2).
\end{equation*}
Over $K$ EM iterations, fitting one MOM model costs
\begin{equation*}
O(KNs^2).
\end{equation*}
Since each sequence is fit from $R$ random initializations, the cost per sequence becomes
\begin{equation*}
O(RKNs^2).
\end{equation*}
Algorithm \ref{Algorithm:2} has two fitting stages: one at state dimension $s$, and one after increasing the state dimension to $s+1$. Therefore, the total MOM fitting cost is
\begin{equation*}
O(J_1RKNs^2 + J_2RKN(s+1)^2).
\end{equation*}
At test time, recursive likelihood scoring of one sequence against $L$ stored MOM models costs
\begin{equation*}
O(LN(s+1)^2),
\end{equation*}
since classification is performed after the state dimension is increased.

For the BPF (shown in Algorithm \ref{Algorithm:3}), let
\begin{itemize}
    \item $N$ be the sequence length (number of time steps)
    \item $P_t$ be the number of particles at time step $t$
    \item $P$ be a representative particle count (e.g., average or typical particle count over time)
    \item $R_{\mathrm{PF}}$ be the number of independent BPF runs (in our implementation, $R_{\mathrm{PF}}=100$)
\end{itemize}

At each time step $t$, the algorithm:
\begin{itemize}
    \item evolves particles independently,
    \item computes weighted summaries and the average weight,
    \item checks which particles branch,
    \item generates offspring for branched particles.
\end{itemize}
Thus, the work at time step $t$ is proportional to
\begin{equation*}
O(P_{t-1}+P_t).
\end{equation*}
Summing over all $N$ time steps, one BPF run has complexity
\begin{equation*}
O\!\left(\sum_{t=1}^N (P_{t-1}+P_t)\right).
\end{equation*}
If the particle count is kept approximately stable (as intended in the branching scheme), i.e.\ $P_t \approx P$, this simplifies to
\begin{equation*}
O(NP).
\end{equation*}
Averaging over $R_{\mathrm{PF}}$ independent runs gives total BPF cost
\begin{equation*}
O(R_{\mathrm{PF}}NP).
\end{equation*}
In our implementation, we used 200 particles and averaged over 100 runs, so this is $O(100NP)$ up to constant factors.

The three methods have different computational profiles, so we compare them separately in terms of fitting/training and test-time scoring. The GAN has a higher up-front optimization cost (epochs, mini-batches, and backpropagation through the generator/discriminator), but test-time classification is cheap once the discriminator is trained (one forward pass, $O(W_D)$ per sequence). In contrast, MOM has heavier fitting cost due to repeated EM iterations and multiple random initializations, and its test-time scoring cost is also larger because each test sequence is compared against a bank of fitted models. The BPF scales linearly in sequence length and particle count, and is the lightest asymptotically per run among the three methods.

\begin{table}[h!]
\centering
\caption{Asymptotic complexity comparison}
\label{asympcomplexitycomparison}
    \begin{tabular}{lcc}
        \hline
        Method & Training / fitting cost & Test-time scoring cost \\
        \hline
        GAN & $O(E(MBW_D + B(W_G+W_D)))$ & $O(W_D)$ per sequence \\
        MOM & $O(J_1RKNs^2 + J_2RKN(s+1)^2)$ & $O(LN(s+1)^2)$ per sequence \\
        BPF & $O(R_{\mathrm{PF}}NP)$ & $O(NP)$ per run/sequence \\
        \hline
    \end{tabular}
\end{table}

This asymptotic comparison is consistent with our implementation-level observation that BPF is computationally cheaper than MOM in our setting, while MOM incurs additional computational cost due to repeated EM fitting and model-bank scoring. In this sense, MOM trades computational time for improved classification performance in our controlled experiments.

\newpage
\section{Algorithms}
\noindent\rule{\linewidth}{0.4pt}
\textbf{Algorithm 1: }GAN Training Process for Coin Flip Sequences\\
\noindent\rule{\linewidth}{0.4pt}
\begin{algorithm}[H]
\DontPrintSemicolon
\KwData{Real coin flip sequences}
\KwInput{Initialize generator $G$ and discriminator $D$ with random weights}
\Fn{\FMainn{$G, D, epochs, batch\_size$}}{
    \For{epoch = 1 to epochs}{

        \textit{\# Train Generator}

        \textit{\# Prepare points in latent space as input for the generator}
        
        $z = \text{generate\_latent\_points}(batch\_size)$
        
        $y_{gan} = 1s$ \textit{\# Create inverted labels for the fake samples}

        \textit{\# Train Discriminator}

        \For{each batch}{
            \textit{\# Get real coin flip sequences}
            
            $X_{real}, y_{real} = \text{generate\_real\_samples}(batch\_size)$
            
            \textit{\# Generate deep fake coin flip sequences}
            
            $z = \text{generate\_latent\_points}(batch\_size)$
            
            $X_{GAN}, y_{GAN} = G(z), 1s$\\

            \textit{\# Get MOM deep fake coin flip sequences $X_{MOM}$ and labels $y_{MOM} = 2$} 
            
            \textit{\# Get handwritten fake coin flip sequences $X_{handfake}$ and labels $y_{handfake} = 3$} 

            \textit{\# Get simulator deep fake coin flip sequences $X_{simulator}$ and labels $y_{simulator} = 4$}
            
            \textit{\# Train discriminator on real and fake samples}
            
            $d\_loss\_real = D.train\_on\_batch(X_{real}, y_{real})$
            
            $d\_loss\_GAN = D.train\_on\_batch(X_{GAN}, y_{GAN})$
            
            $d\_loss\_MOM = D.train\_on\_batch(X_{MOM}, y_{MOM})$
            
            $d\_loss\_handfake = D.train\_on\_batch(X_{handfake}, y_{handfake})$
            
            $d\_loss\_simulator = D.train\_on\_batch(X_{simulator}, y_{simulator})$
        }
        
        \textit{\# Update the generator via the discriminator's error}
        
        $g\_loss = GAN.train\_on\_batch(z, y_{gan})$
        
        \textit{\# Print the progress}
        
        \If{epoch \% 100 == 0}{
            Print "Epoch", epoch, 
            
            Print "Discriminator Loss:", 
            (d\_loss\_real + d\_loss\_GAN + d\_loss\_MOM + d\_loss\_handfake + d\_loss\_simulator) / 5, 
            
            Print "Generator Loss:", g\_loss
        }
    }
}
\noindent\rule{\linewidth}{0.4pt}

\label{Algorithm:1}
\end{algorithm}

\newpage

\noindent\rule{\linewidth}{0.4pt}
\textbf{Algorithm 2: }MOM Training Process\\
\noindent\rule{\linewidth}{0.4pt}

\begin{algorithm}[H]
\DontPrintSemicolon
  \KwData{Observations of real and fake coin sequences, $Y$}
  \KwInput{Initialize transition probabilities $p_{x \rightarrow x'}, q_{y \rightarrow y'(x)}$ and initial distribution $\mu(x,y)$}
  \Fn{\FMain{$p's,q's,\mu's,N,Y$}}{
    \For{each real training sequence $i$}{
     \textit{\# forward propagation}\\
     $\pi_0 (x,y) = \mu(x,y)$      
     $\pi_1(x) =$ $\frac{\sum\limits_{x_0 \in E} \sum\limits_{y_0 \in O} \mu(x_0 , y_0)p_{x_0\rightarrow x}q_{y_0\rightarrow Y_1}}{\sum\limits_x (\sum\limits_{x_0 \in E} \sum\limits_{y_0 \in O} \mu(x_0 , y_0)p_{x_0\rightarrow x}q_{y_0\rightarrow Y_1})}$\\
\For{n = 2, 3, ..., N}{
     $\pi_n (x) = \frac{q_{Y_{n-1}\rightarrow Y_n}(x) \sum_{x_{n-1}} \pi_{n-1}(x_{n-1}) p_{x_{n-1}\rightarrow x}}{\sum_x q_{Y_{n-1}\rightarrow Y_n}(x) \sum_{x_{n-1}} \pi_{n-1}(x_{n-1}) p_{x_{n-1}\rightarrow x}}$
}

    \textit{\# backward propagation}\\
      $\chi_{N-1}(x)=q_{Y_{N-1}\rightarrow Y_n}(x)$\\
       $\hspace{0.5cm}$$\chi_0 (x,y) = \frac{q_{y\rightarrow Y_1}(x)}{\sum\limits_x (\sum\limits_{x_0 \in E} \sum\limits_{y_0 \in O} \mu(x_0 , y_0)p_{x_0\rightarrow x}q_{y_0\rightarrow Y_1})}$\\
       
    \For{n = N-1,N-3,...,1}{
       $\chi_n (x) = \frac{q_{Y_n \rightarrow Y_{n+1}}(x)}{\sum\limits_x q_{Y_n \rightarrow Y_{n+1}}(x) \sum\limits_{x_{n}} \pi_n (x_n)p_{x_{n}\rightarrowx}}$
    }  
    \textbf{Compute} $p_{i_{\text{real}}}$, $q_{i_{\text{real}}}$, $\mu_{i_{\text{real}}}$ in state $s$ via EM
            
               $p_{x \rightarrow x’} =$    $\frac{ p_{x\rightarrow x’}\big[ \sum\limits_y \pi_0 (x,y)\chi_0 (x’,y)+ \sum\limits^{N-1}_{n=1} \pi_n (x) \chi_n (x’) \big]}{\sum\limits_{x_1} p_{x\rightarrow x_1} \big[ \sum\limits_y \pi_0 (x,y) \chi_0(x_1,y)+ \sum\limits^{N-1}_{n=1} \pi_n (x) \chi_n (x_1) \big]}$ 

              $q_{y \rightarrow y’}(x) = \frac{\sum\limits_\xi p_{\xi \rightarrow x} \big[ 1_{Y_{1=y’}}\chi_0(x,y) \pi_0 (\xi,y)+ \sum\limits^{N-1}_{n=1}1_{Y_{n=y},Y_{n+1 =y’}} \chi_n (x) \pi_n (\xi) \big]}{ \sum\limits_\xi p_{\xi \rightarrow x} \big[ \chi_0(x,y) \pi_0 (\xi,y)+ \sum\limits^{N-1}_{n=1}1_{Y_{n=y}} \chi_n (x) \pi_n (\xi) \big]}$  

              $\mu (x,y) =\frac{\mu(x,y) \sum\limits_{x_1} \chi_0 (x_1,y) p_{x \rightarrow x_1} }{\sum\limits_\xi \sum\limits_\theta \mu (\xi,\theta) \sum\limits_{x_1}  \chi_0 (x_1 ,\theta) p_{\xi \rightarrow x_1}}$

    \textbf{Simulate} coin sequence $j$ generated using $p_{i_{\text{real}}}$, $q_{i_{\text{real}}}$, $\mu_{i_{\text{real}}}$

    \For{each model $p_{i_{\text{real}}}, q_{i_{\text{real}}}, \mu_{i_{\text{real}}}$}{
              Compute recursive likelihood between input sequence and model
        }
          \textit{\# Compare recursive likelihoods to classify the sequence}\\
          \textit{\# Assign labels based on highest recursive likelihood}
    }
    \textbf{Raise} state dimension to $s + 1$ from state $s$\\
    
    \For{each real training sequence $i$}{
          \textbf{Re-compute} Final estimates of  $p_{i_{\text{real}}}$, $q_{i_{\text{real}}}$, $\mu_{i_{\text{real}}}$ in state $s+1$ via EM
    }
    \For{other sequences (MOM, GAN, handwritten, simulator)}{
          \textbf{Compute} recursive likelihoods and estimate $p$, $q$, $\mu$ in state $s+1$ via EM
    }

   }

\noindent\rule{\linewidth}{0.4pt}

\label{Algorithm:2}
\end{algorithm}

\newpage
\noindent\rule{\linewidth}{0.4pt}
\textbf{Algorithm 3: }Branching Algorithm\\
\noindent\rule{\linewidth}{0.4pt}
\begin{algorithm}[H]
\DontPrintSemicolon
  \Fn{\FTest{$X_0,L_0 N,T,r$}}{
    \For{$t = 1$ to $T$}{
           \textit{Evolve particles independently}
	     $(X^{j}_{t-1}, L^{j}_{t-1}) = (\hat{X}^{j}_{t}, \hat{L}^{j}_{t})$
   
           Estimate $\sigma_t$ by $S^N_t = \frac{1}{N} \sum\limits^{N_{t-1}}_{j=1}\hat{L}^j_t  \delta_{\hat{X}^j_t}$
        
	      \textit{Calculate average weight}
	     $A_t = S^N_t (1)$
   
	     \textit{Check which particles to branch}
   
	     $m = 0$
   
	  \For{$j = 1$ to $N_{t-1}$}{
		  \If{$\hat{L}^j_t \in \big( \frac{1}{r_t}A_t, r_t A_t \big)$}{
               \textit{Move non-branched particles to final vector}
                   $(X^{j-m}_{t}, L^{j-m}_{t}) = (\hat{X}^{j}_{t}, \hat{L}^{j}_{t})$
            }    
          \Else{
                   $m = m+1$
                
                   $(X^{m}_{t}, L^{m}_{t}) = (\hat{X}^{j}_{t}, \hat{L}^{j}_{t})$
               }
           
        }
           \textit{Branching part of the algorithm}
           $N_t = m$
        
	     Simulate $\{V^j_t\}^{N_{t-1}}_{j = m+1}$ with $V^j_t \sim \big[ \frac{j-m-1}{N_{t-1}-m},\frac{j-m}{N_{t-1}-m} \big]$-Uniform
   
           Let $p$ be a random permutation of $\{m+1, m+2 \dots N_{t-1} \}$
        
           $U^j_t = V^{p(j)}_{t}$
        
        \For{$j = m+1$ to $N_{t-1}$}{
               $N^j_t = \lfloor \frac{\hat{L}^{j-l}_t}{A_t}\rfloor + 1_{\big\{ U^j_t \leq \big( \frac{\hat{L}^{j-l}_t}{A_t} - \lfloor \frac{\hat{L}^{j-l}_t}{A_t} \rfloor \big) \big\} } $
            
                \For{$k=1$ to $N^j_t$}{
                       $(X^{N_t +k}_{t}, L^{N_t +k}_{t}) = (\hat{X}^{j-l}_{t}, A_t)$
                }
                   $N_t = N_t + N^j_t$
        } 
    }
        
    }

\noindent\rule{\linewidth}{0.4pt}

\label{Algorithm:3}
\end{algorithm}

\end{document}